\documentclass[letterpaper, 10 pt, conference]{ieeeconf}  

\IEEEoverridecommandlockouts                              

\usepackage{amsfonts}
\usepackage{amsmath}
\newtheorem{definition}{Definition}
\usepackage{algorithm}
\usepackage{algorithmic}
\newcommand{\veryshortarrow}[1][3pt]{\mathrel{%
   \hbox{\rule[\dimexpr\fontdimen22\textfont2-.2pt\relax]{#1}{.4pt}}%
   \mkern-4mu\hbox{\usefont{U}{lasy}{m}{n}\symbol{41}}}}
\usepackage{subfigure}
\usepackage{subcaption}
\usepackage{graphicx}
\usepackage{adjustbox}
\usepackage{booktabs} 
\usepackage{todonotes}
\title{\LARGE \bf
Estimating Control Barriers from Offline Data\thanks{This paper has been accepted to ICRA 2025.}
}

\author{Hongzhan Yu$^{1}$, Seth Farrell$^{1}$, Ryo Yoshimitsu$^{2}$, Zhizhen Qin$^{1}$, Henrik I. Christensen$^{1}$ and Sicun Gao$^{1}$
    \thanks{$^{1}$ Affiliation: 1. UCSD, 2. IHI Japan. Email: {\tt\small hoy021@ucsd.edu}}%
}

\begin{document}

\maketitle
\thispagestyle{empty}
\pagestyle{empty}

\begin{abstract}
Learning-based methods for constructing control barrier functions (CBFs) are gaining popularity for ensuring safe robot control.
A major limitation of existing methods is their reliance on extensive sampling over the state space
or online system interaction in simulation.
In this work we propose a novel framework for learning neural CBFs through a fixed, sparsely-labeled dataset collected prior to training.
Our approach introduces new annotation techniques based on out-of-distribution analysis, enabling efficient knowledge propagation from the limited labeled data to the unlabeled data. 
We also eliminate the dependency on a high-performance expert controller, and allow multiple sub-optimal policies or even manual control during data collection.
We evaluate the proposed method on real-world platforms.
With limited amount of offline data, it achieves state-of-the-art performance for dynamic obstacle avoidance, demonstrating statistically safer and less conservative maneuvers compared to existing methods. 
\end{abstract}

\section{Introduction}

Control Barrier Functions (CBFs) provide an effective framework for safe robot control~\cite{ames2014control,ames2016control}.The recent development of learning-based CBF methods exploit the expressiveness of neural networks and data-driven approaches to handle systems with complex dynamics and high uncertainty, with promising results~\cite{wang2018safe,taylor2020learning,qin2021learning,dawson2022safe,saveriano2019learning, robey2020learning,zhao2021learning}. 

However, the scalability of learning-based methods has been a major bottleneck. 
The typical approach for learning neural CBFs requires sampling over the entire state space to enforce constraints from the standard CBF conditions~\cite{qin2022quantifying}.
While such methods ensure comprehensive coverage, they are impractical for high-dimensional domains due to the need of exponentially many samples. 
Online methods~\cite{hoi2021online} have been proposed to mitigate this issue by interacting with the system during learning, allowing the state space to be gradually explored.
However, the methods still rely on the availability of high-fidelity simulations and online interactions for learning at arbitrary system states, and can not operate with a fixed pre-collected dataset. 

\begin{figure}[t]
    \centering
    \adjustbox{width=0.45\textwidth}{
        \includegraphics{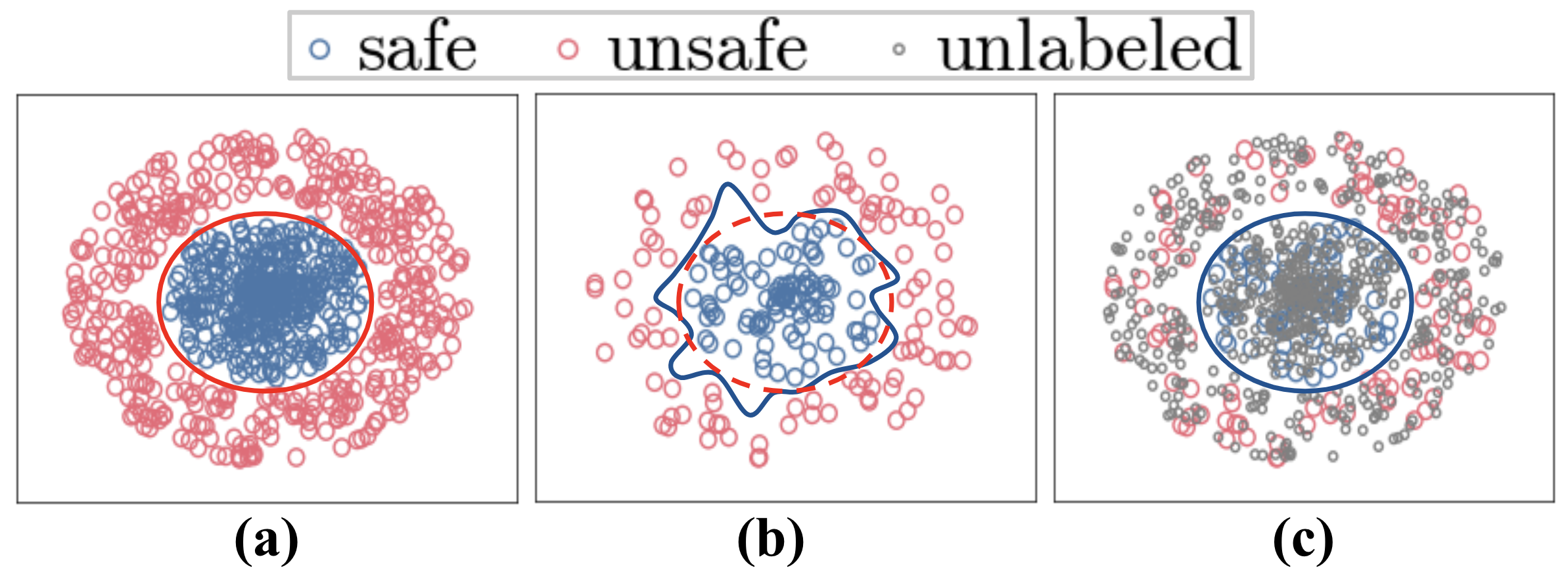}
    }
    \caption{\small Visualizations on toy datasets, to illustrate the motivation for utilizing unlabeled data.
    \textbf{(a)} With sufficient labeled data, the model can accurately capture the safety boundary. 
    \textbf{(b)} When labeled data is limited, the learned boundary often misclassifies the safe and unsafe regions of the system.
    \textbf{(c)} Unlabeled data is generally more accessible than labeled data. 
    Our approach leverages unlabeled data, along with the limited labeled data, to capture the CBF landscape that best adhere to the constraints inherent in the data.
    }
    \label{fig:unlabeled_motivation}
\end{figure}

Consequently, {\em offline} training for constructing CBFs from a pre-collected dataset is an important and mostly open problem. 
Existing  methods that consider the offline setting impose restrictive assumptions on the training data, such as only utilizing successful trajectories~\cite{saveriano2019learning, robey2020learning, qin2022quantifying},
or relying on a fixed performative controller for data collection~\cite{robey2020learning, yu2023sequential, sun2021learning, xiao2023barriernet}.
In contrast, leveraging unlabeled trajectories collected by untrained controllers, which is the more realistic setting of data collection, 
is key to achieving offline learning in practice. 
Thus, the core question is how to harness a small set of labeled data to train CBFs from a larger set of unlabeled suboptimally-collected behavior data of the robot (Figure \ref{fig:unlabeled_motivation}). 
Note that given the learning-based setting of the problem, our focus is not to derive complete guarantee of safety, but to achieve better control performance using CBFs compared to existing methods.

\begin{figure}[t]
    \centering
    \adjustbox{width=0.49\textwidth}{
        \includegraphics{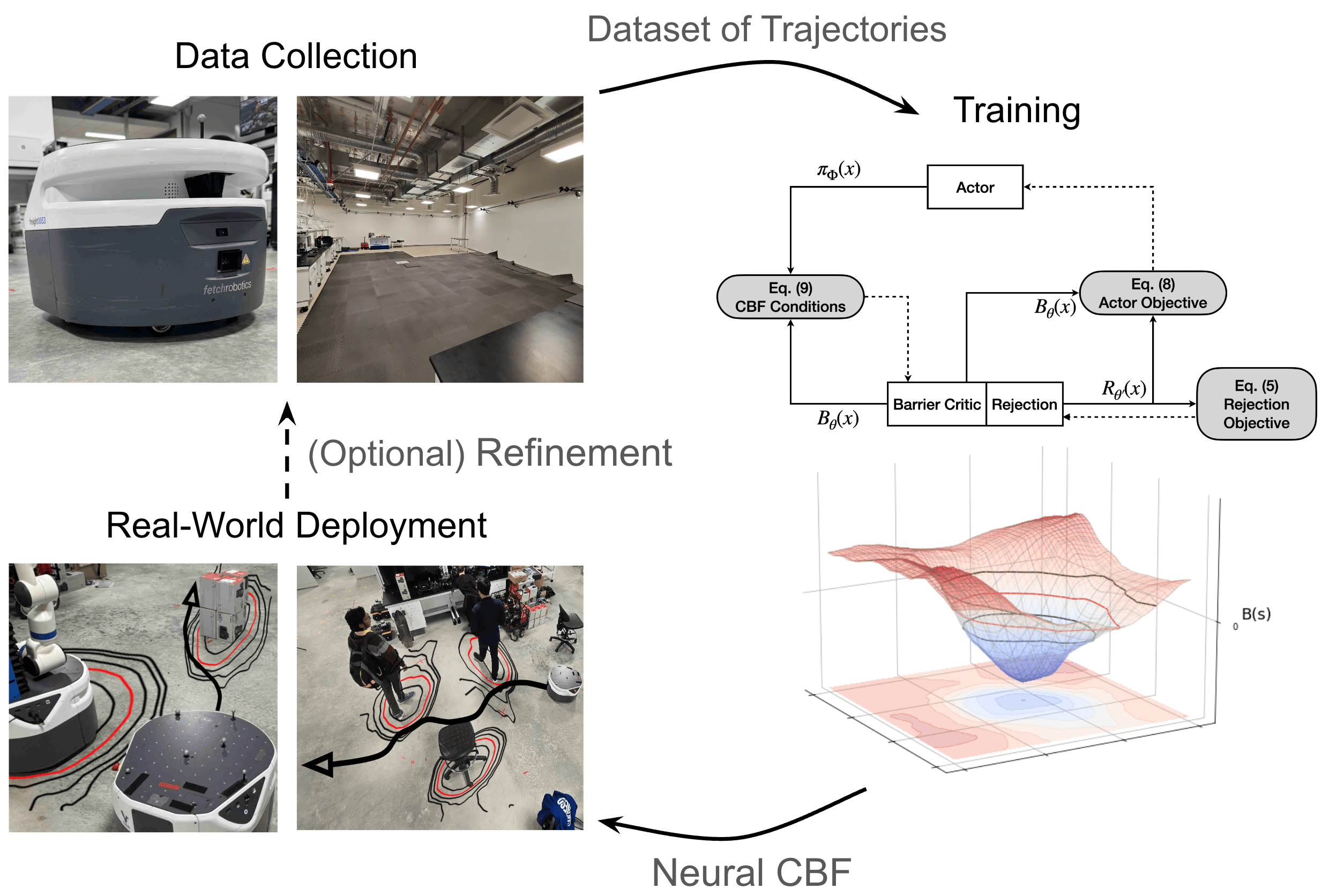}
    }
    \caption{\small
    Overall learning pipeline of the proposed method.
    Our method utilize offline demonstrations 
    - even directly collected from real-world platforms -
    to construct neural control barriers,
    ensuring their zero-superlevel sets are control-invariant.
    The red contours denote the learned zero level set, serving as the safety boundary the ego-robot must not cross.
    Optionally, 
    we gather additional demonstrations to further refine the barrier.
    } \label{fig:overall}
\end{figure}

Overall, we introduce a novel offline learning framework for constructing neural CBFs (Figure \ref{fig:overall}).
First, we propose new learning procedures that leverage out-of-distribution analysis~\cite{yang2024generalized} of trained neural models to propagate insights from labeled data, thereby maximizing the utility of unlabeled data.
Second, we remove the reliance on high-performance controllers during data collection, enabling sub-optimal controllers to gather data while still achieving successful offline training.
To achieve this, we improve the learning of CBFs to more closely align with their theoretical Lie-derivative condition. 
We evaluate the proposed method for obstacle avoidance with autonomous mobile robots on both simulation and real-world platforms. With limited amount of offline data, the proposed methods can achieve state-of-the-art performance in dynamic obstacle avoidance.
The paper is organized as follows.
Sections \ref{section:related} and \ref{section:preliminary} cover the related work and the background necessary to support our theory.
In Section \ref{section:algorithm}, we introduce the proposed algorithm and explain the rationale behind each component.
Section \ref{section:experiment} presents the experimental results,
and Section \ref{section:conclusion} concludes the paper.


\section{Related Work}
\label{section:related}
\textbf{Control Barrier Function.} 
Control barrier functions (CBF) \cite{ames2019control} aim to ensure control safety in dynamical systems by imposing value-landscapes to render the safe set forward invariant. 
The key point is to enforce the derivative of the CBF to satisfy Lyapunov conditions~\cite{lyapunov1992general}.
Traditional CBFs are manually designed based on domain-specific knowledge of the system, making them unsuitable for systems with complex dynamics or high uncertainty~\cite{ames2014control, ames2016control}.

Learning-based methods have been introduced to construct data-driven CBF candidates~\cite{{wang2018safe, taylor2020learning,qin2021learning,dawson2022safe,robey2020learning,yu2023sequential,zhang2023neural,yu2023learning}}
from data.
Online algorithms learn CBFs by interacting with, or sampling from, the controlled system.
In \cite{wang2018safe}, the authors learn barrier certificates to derive the safe region of an unknown control-affine system. 
They propose an adaptive sampling algorithm to iteratively refine the CBF candidate on the states that have high uncertainty.
\cite{qin2021learning} studies the multi-agent control problem. They jointly learn the barrier certificates alongside the multi-agent control policy, while regulating the policy based on CBF.
\cite{dawson2022safe} develops a model-based approach to learn control Lyapunov barrier functions based on stability and safety specifications.
The training state are sampled uniformly from the state space.
Offline algorithms learn CBFs without new data during the learning.
\cite{saveriano2019learning} proposes an incremental learning of a set of linear parametric CBFs from human demonstrations.
In \cite{robey2020learning}, the authors present an approach to synthesize local valid CBFs for control-affine systems with known but nonlinear dynamics.
The expert demonstrations contain only safe trajectories collected with a fixed nominal controller.

\textbf{Out-of-distribution Analysis.}
Out-of-distribution (OOD) analysis is an emerging topic of machine learning that examines the distribution shifts where test data diverges from the training data distribution \cite{yang2024generalized}.
Unsupervised representation learning methods focus on learning domain-agnostic features from unlabeled data~\cite{mahajan2021domain,zhang2022towards,harary2022unsupervised,charoenphakdee2021classification}.
However, these methods can introduce bias, if the OOD domain distributions overlap with the unlabeled data distribution~\cite{yu2024rethinking}.
Supervised learning methods incorporate implicit domain labels from both in-distribution and OOD data~\cite{arjovsky2019invariant, wu2022discovering}.
While these methods are often more accurate due to the additional information, they may not generalize well to OOD examples that differ significantly from those seen during training.

\section{Preliminary}
\label{section:preliminary}

We consider ego-robots with underlying dynamics $\dot{x}(t) = f(x(t), u(t))$ where $x(t)$ takes values in an $n$-dimensional state space $\mathcal{X} \subseteq \mathbb{R}^n$, $u(t) \in \mathcal{U} \subseteq \mathbb{R}^{m}$ is the control vector, and $f: \mathcal{X} \times \mathcal{U} \rightarrow \mathcal{X}$ is a Lipschitz-continuous vector field.
We allow the dynamics function $f$ to be generally nonlinear and not control-affine.
Consider an unsafe region of the state space $\mathcal{X}_{u} \subset \mathcal{X}$ where safety constraints are violated. We say the system is safe if none of its trajectories intersects with $\mathcal{X}_{u}$. 
To enforce safety properties of a system, the controller needs to find a \textit{control invariant} set for the system that is disjoint from the unsafe set. 
A subset of the state space $\mathsf{Inv} \subseteq \mathcal{X}$ is control invariant, if for any initial state $x(0) \in \mathsf{Inv}$ and any $t > 0$, we have $x(t) \in \mathsf{Inv}$. 
Namely, any trajectory that starts in the invariant set $\mathsf{Inv}$ stays in $\mathsf{Inv}$ forever.
CBFs are scalar functions whose zero-superlevel set is a control invariant set within the safe region of the system, and whose spatial gradients can be used to enforce the invariance.

\begin{definition}[Control Barrier Functions~\cite{ames2019control}] \label{def:barrier_func}
Consider a dynamical system defined by vector field $f:\mathcal{X} \times \mathcal{U} \rightarrow \mathcal{X}$. 
Let $B:$ $\mathcal{X}$ $\rightarrow$ $\mathbb{R}$ be a continuously differentiable function.
The Lie derivative of $B$ over $f$ is defined as:
\begin{equation}
\begin{aligned}
    L_{f,u} B(x) &=  \sum_{i=1}^{n} \frac{\partial B}{\partial x_{i}} \cdot \frac{\partial x_{i}}{\partial t} 
    = \langle \nabla_{x} B(x), f(x, u)\rangle,
\end{aligned}
\end{equation}
where $\langle \cdot,\cdot \rangle$ denotes inner product.
The Lie derivative measures the change of $B$ over time along the direction of system dynamics under control $u$.
If the zero-superlevel set of $B$, i.e. $\mathcal{C}=\{x\in \mathcal{X}: B(x)\geq 0\}$, is disjoint from the unsafe region of the system, i.e. $\mathcal{C} \cap \mathcal{X}_{u} = \emptyset$.
And if for any safe state $x \in \mathcal{C}$ and an extended class-$\mathcal{K}_{\infty}$ function ${\alpha} (\cdot)$~\cite{Khalil:1173048}:
\begin{equation} \label{equ:barrier_function}
\begin{aligned}
\max_{u\in \mathcal{U}} L_{f,u} B(x) \geq -\alpha (B(x)).
\end{aligned}
\end{equation}
Then $B$ is a control barrier function (CBF), and its zero-superlevel set $\mathcal{C}$ is control invariant.
\end{definition}

Out-of-distribution (OOD) algorithms study whether an input to the neural model follows the training distribution, being \textit{in-distribution}, or deviates from the model's training set, making it \textit{out-of-distribution}.
In our work, we use the unsupervised algorithm \cite{charoenphakdee2021classification} to implement OOD detection.
Consider the input space $\mathcal{X}$ and the binary output space $\mathcal{Y} = \{-1, +1\}$.
For a threshold $c \in (0, 1) \subseteq \mathbb{R}$, a binary classifier $P: \mathcal{X} \rightarrow (0, 1)$ is optimized to capture the correct labels, where the classification decision $f_{c, P}(x)$ is set to $+1$ only if $P(x) > c$.
To achieve binary classification with rejection, \cite{charoenphakdee2021classification} proposes learning two binary classifiers, denoted as $P_{1}$ and $P_{2}$, with thresholds $c$ and $1-c$, respectively, to satisfy Chow's rule~\cite{chow1970optimum}. 
The two models share all the model weights except for the last layer.
If the two classifiers disagree on the classification of a given input $x$, i.e. $f_{c, P_{1}}(x) \neq f_{1-c, P_{2}}(x)$, the input is rejected as OOD. 

\section{Offline Learning of Barrier Critic}
\label{section:algorithm}



Consider that we have a well-defined CBF $B: \mathcal{X} \rightarrow \mathbb{R}$, and a discrete control system $f: \mathcal{X} \times \mathcal{U} \rightarrow \mathcal{X}$.
For an arbitrary unlabeled state $o \in \mathcal{X}$ that does not violate safety, i.e. $o \notin \mathcal{X}_{u}$, 
if there exist controls at $o$ that can lead the system to be in the zero-superlevel set of $B$:
\begin{equation} \label{eqn:annotating}
\begin{aligned}
    \exists u \in \mathcal{U} \text{ } s.t. \text{ } B(f(o, u)) > 0,
\end{aligned}
\end{equation}
then the state $o$ satisfies the control invariant property, and thus, assigning a safe label to it must be correct.
However, for the data-driven neural CBF models,
following (\ref{eqn:annotating}) can lead to incorrect annotations of the unlabeled.
This is because if an unlabeled state $o$ is uncovered by the training set, it is likely that neither is its one-step reachable set, i.e. $\mathcal{X}^{'} = \{x \in \mathcal{X} \text{ } | \text{ } \exists u \in \mathcal{U} \text{ } s.t. \text{ } f(o, u) = x\}$, covered fully.
Thus, the model predictions on $\mathcal{X}^{'}$ are not reliable in determining the safety of $o$.
This phenomenon is critical for offline methods, as we cannot interact with the system to acquire on-policy data to refine on the wrongly-annotated state regions.

In our work, we propose to label an unlabeled state $o$ as safe if there exists a control $u \in \mathcal{U}$ such that:
\begin{equation} \label{eqn:annotating2}
\begin{aligned}
    B_{\theta}(x')> 0  \textit{\textbf{ and }} x' \text{ is } \textit{in-distribution} \text{ w.r.t. $\theta$},
\end{aligned}
\end{equation}
where $x' = f(o, u)$ and $\theta$ represents the parameters of neural CBF model.
Intuitively, if we can derive such a control that leads the system to a \textit{seen \& safe} state, then there arises no concern about undermining the annotation steps due to the OOD samples.
In the following sections, we describe the components for achieving the proposed idea. 

\subsection{Rejection-based Out-of-distribution Analysis}\label{section:rejection}

We employ \cite{charoenphakdee2021classification} to determine whether a given input is in-distribution.
Let the rejection model be denoted by $R_{\Phi}: X \rightarrow \mathbb{R}^{2}$, which outputs two-dimensional rejection scores for the given state input.
Denote the rejection threshold by $c \in (0, 1) \subseteq \mathbb{R}$.
Over the safe set $\mathbb{X}_{s}$ and the unsafe set $\mathbb{X}_{u}$, we optimize $R_{\Phi}$ to minimize the following objective:
\begin{align}\label{eqn:rejection_objective}
\begin{split}    
    L_{\Phi}(\mathbb{X}_{s}, \mathbb{X}_{u}) &= L_{{\Phi_{1}}, c}(\mathbb{X}_{s}, \mathbb{X}_{u}) + L_{{\Phi_{2}}, 1-c}(\mathbb{X}_{s}, \mathbb{X}_{u}), \\ 
    L_{{\Phi}_{i}, (\cdot)}(\mathbb{X}_{s}, \mathbb{X}_{u}) &= \frac{1}{|\mathbb{X}_{s}|} \sum_{x \in \mathbb{X}_{s}} [- R_{\Phi_{i}}(x) + (\cdot)]_{+} \\ 
    & + \frac{1}{|\mathbb{X}_{u}|} \sum_{x \in \mathbb{X}_{u}} [ R_{\Phi_{i}}(x) - (\cdot)]_{+}, 
\end{split}
\end{align}
where $R_{\Phi_{i}}(x)$ denotes the $i^{th}$ score from $R_{\Phi}(x)$ and $[\cdot]_{+} = \max(\cdot, 0)$.
When the rejection model is well trained, we say that the state $x \in \mathcal{X}$ is an in-distribution sample if:
\begin{equation} \label{eqn:rejection_condition2}
    R_{\Phi_{1}}(x) > c \textit{\textbf{ and }} R_{\Phi_{2}}(x) > 1 - c,
\end{equation}
implying no disagreement between the two rejection scores.

\subsection{Actor Model Learning}\label{section:actor}

The rejection model enables us to classify if a given state is in-distribution or not.
However, to realize the annotation steps proposed in (\ref{eqn:annotating2}), we must
be able to efficiently determine what controls to attempt at one unlabeled state.

We achieve this by learning an actor model $A_{\Theta}: \mathcal{X} \rightarrow \mathcal{U}$ that captures the maximally-safe, in-distribution control for the given state.
The term `maximally-safe' is with respect to the CBF landscape, accounting for the maximal increase to the learned CBF score led by the control.
Denote the CBF model by $B_{\theta}: \mathcal{X} \rightarrow \mathbb{R}$.
With the rejection model $R_{\Phi}$ and the parameter $c$, we aim at solving the following optimization problem with the actor at an arbitrary state $x \in \mathcal{X}$:
\begin{align}\label{eqn:actor_goal}
\begin{split}
    \arg\max_{u \in \mathcal{U}} \text{ }&B_{\theta}(f(x, u)), \\ 
    \textit{s.t. } R_{\Phi_{1}}(f(x, u)) > c &\textit{ and } R_{\Phi_{2}}(f(x, u)) > 1 - c.
\end{split}
\end{align}
Consider that we obtain the control $u^{*}$ by solving (\ref{eqn:actor_goal}) at an unlabeled state $o$.
If following $u^{*}$ at $o$ cannot satisfy (\ref{eqn:annotating2}), then no less safe control can satisfy it either.
Therefore, we can label $o$ as unsafe without evaluating any other controls.
Given a training batch $\mathbb{X}$, we optimize $\Theta$ by minimizing:
\begin{align} \label{eqn:actor_loss}
\begin{split}
L_{\Theta}(\mathbb{X}) &= \frac{1}{|\mathbb{X}|} \sum_{x \in \mathbb{X}} \Bigg[ -B_{\theta}\Big(f(x, \pi_{\Theta}\big(x\big))\Big) \\ 
&+ [- R_{\Phi_{1}}\Big(f(x, \pi_{\Theta}\big(x\big))\Big) + c]_{+}\\ 
&+ [- R_{\Phi_{2}}\Big(f(x, \pi_{\Theta}\big(x\big))\Big)+1-c]_{+}\Bigg].
\end{split}
\end{align}
Unlike Reinforcement Learning (RL) methods~\cite{lillicrap2015continuous, fujimoto2018addressing}, we do not rely on the actor to generate controls at execution time. Instead, the actor is used soly as an auxiliary model to shape the barrier landscape during training.

\begin{figure}[t]
    \centering
    \adjustbox{width=0.45\textwidth}{
        \includegraphics{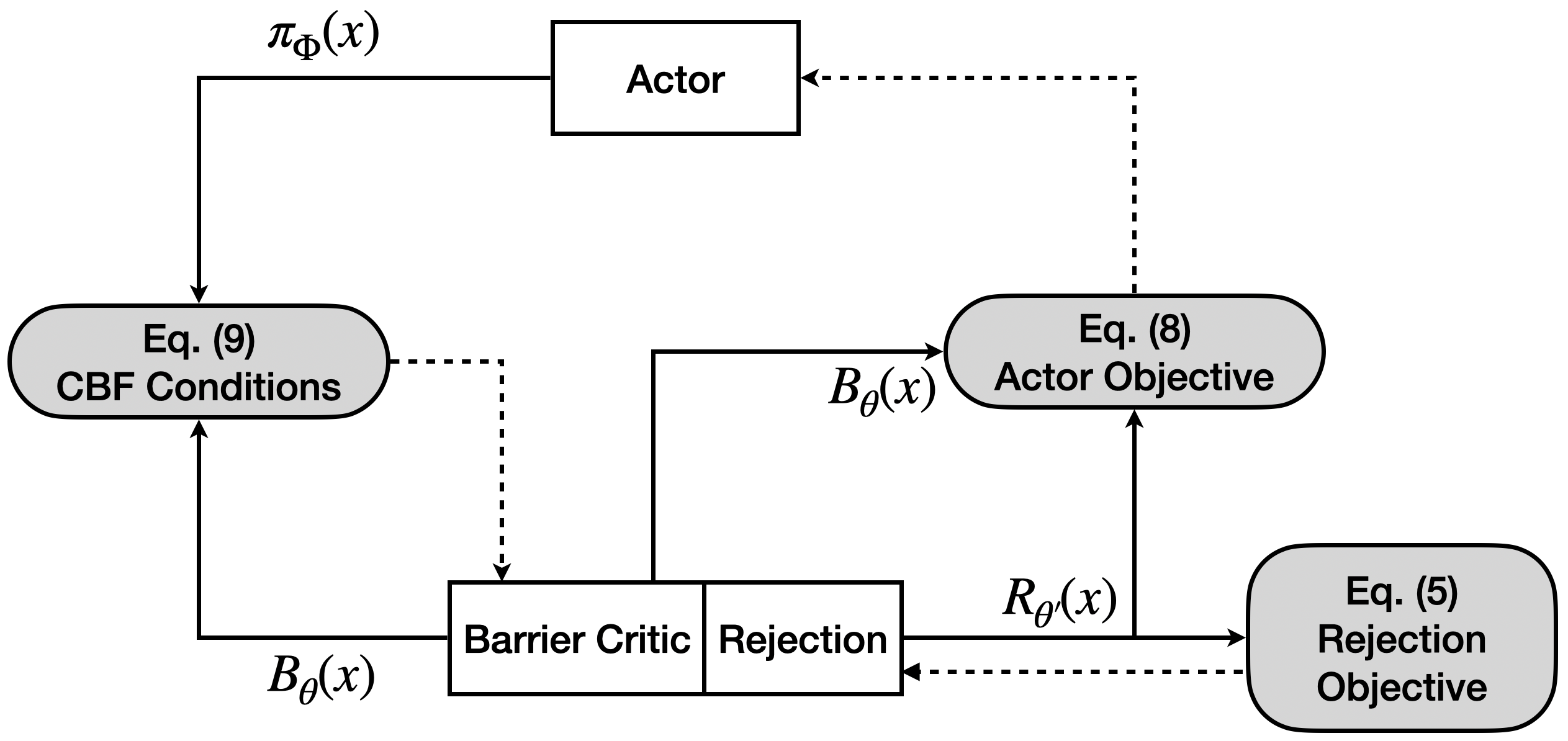}
    }
    \caption{\small Model component visualization with the proposed training flows.
    The optimization of the CBF
    requires the actor to derive the maximally-safe controls over which we enforce Lie derivative condition. 
    The actor is optimized based on both CBF and rejection models, capturing the control that leads to the \textit{safest in-distribution} state.
    Rejection model's training does not rely on other models.
    }
    \label{fig:model_component}
\end{figure}

\begin{figure}[t]
    \centering
    \adjustbox{width=0.45\textwidth}{
        \includegraphics{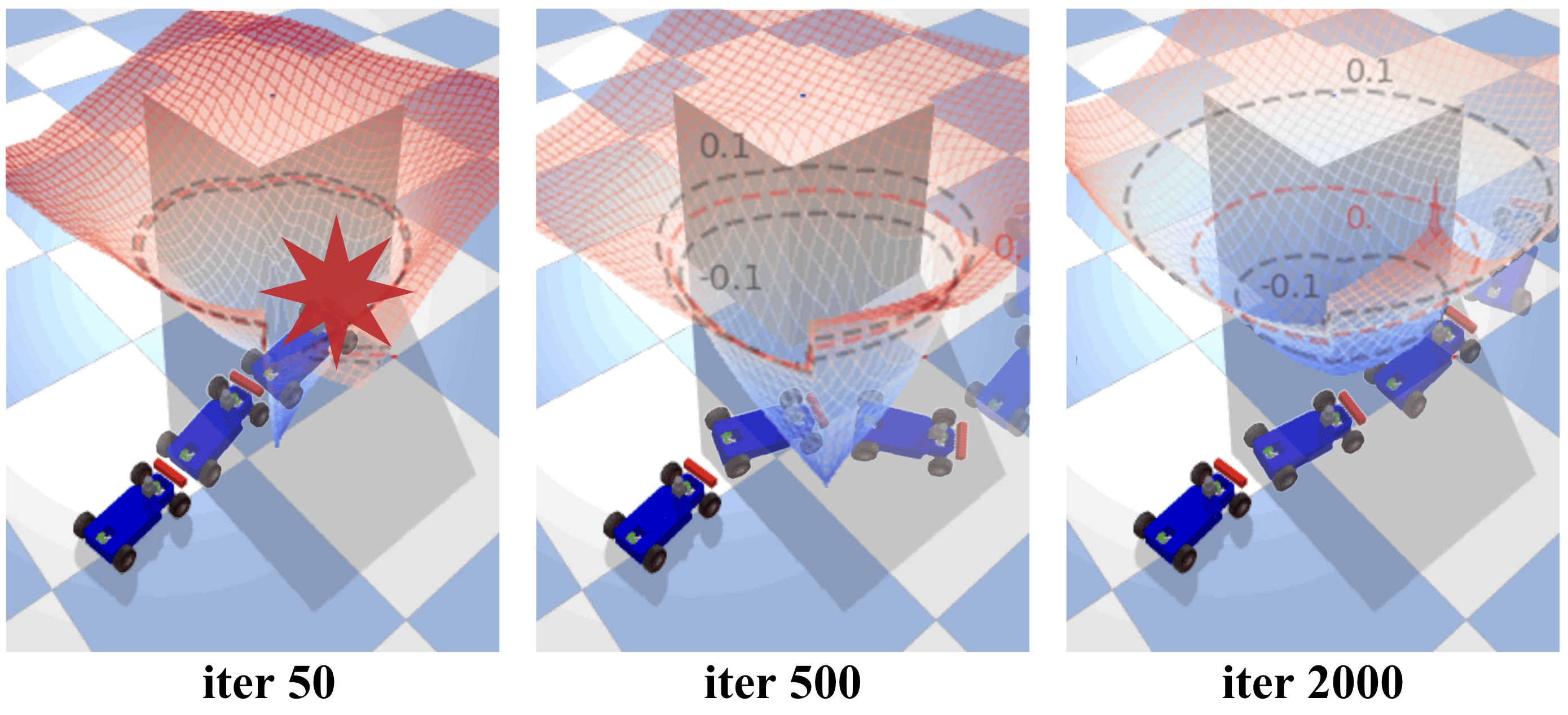}
    }
    \caption{\small Visualization of the learned CBF landscapes.
    Early in training, the safety boundary is primarily informed by the collision states from data,
    leading the robot to collide.
    As training progresses, the CBF model begins to approximate the safe region of the system.
    However, jittery robot motions are exhibited due to incomplete training.
    Once learning converges, the CBF model satisfies the Lyapunov condition over its Lie-derivatives, enabling the selection of more aggressive yet still safe controls.
    }
    \label{fig:simulation_visual}
\end{figure}

\subsection{Overall Pipeline}\label{section:pipeline}

\begin{algorithm}[t]
 \caption{Neural CBF with Barrier Critic (NCBF-BC)} \label{full_alg}
 \begin{algorithmic}[1]
 \renewcommand{\algorithmicrequire}{\textbf{Input:}}
 \REQUIRE labeled sets $D_{s}$ and $D_{u}$, unlabeled set $D_{ul}$, training iteration $T$, annotation start iteration $T_{a}$\\
 \STATE \textit{Initialize} the models and data buffer
 \FOR {$t = 1$ ... $T$}
   \STATE Sample labeled batches $\mathbb{X}_{s} \subseteq D_{s}$ and $\mathbb{X}_{u} \subseteq D_{u}$
   \IF {($t \geq T_{a}$)} \label{algline:annotate_condition}
   \STATE Sample an unlabeled batch $\mathbb{X}_{ul} \subseteq D_{ul}$
   \STATE $\mathbb{X}_{ul, s}$, $\mathbb{X}_{ul, u}$ $\gets$ \textit{Annotate}($\mathbb{X}_{ul}$) \label{algline:annotate}
   \STATE $\mathbb{X}_{s} \gets \mathbb{X}_{s} \cup \mathbb{X}_{ul, s}$, $\mathbb{X}_{u} \gets \mathbb{X}_{u} \cup \mathbb{X}_{ul, u}$
   \ENDIF
    \STATE Model \textit{updates} over $\mathbb{X}_{s}$ and $\mathbb{X}_{u}$: $R_{\Phi}$ with (\ref{eqn:rejection_objective}), $\pi_{\Theta}$ with (\ref{eqn:actor_loss}), $B_{\theta}$ with (\ref{eqn:barrier_objective})
 \ENDFOR
 \RETURN $B_{\theta}$, $R_{\Phi}$, $\pi_{\Theta}$
 \\ \text{}
 \\\textbf{Function} \textit{Annotate}($\mathbb{X}$):
 \STATE \hspace{\algorithmicindent}$\mathbb{X}_{s} \gets \{\}$, $\mathbb{X}_{u} \gets \{\}$
 \STATE \hspace{\algorithmicindent}\textbf{for} $x$ in $\mathbb{X}$ \textbf{do}
 \STATE \hspace{\algorithmicindent}\hspace{\algorithmicindent}$\bar{x} \gets f(x, \pi_{\Theta}(x))$  \label{algline:annotation_unroll}
 \STATE \hspace{\algorithmicindent}\hspace{\algorithmicindent}\textbf{if} {$B_{\theta}(\bar{x}) > 0$ \textbf{and} $R_{\Phi}(\bar{x})$ satisfies (\ref{eqn:rejection_condition2})} \textbf{then} \label{algline:annotation_rule}
 \STATE \hspace{\algorithmicindent}\hspace{\algorithmicindent} \hspace{\algorithmicindent}$\mathbb{X}_{s} \gets \mathbb{X}_{s} \cup \{x\}$
 \STATE \hspace{\algorithmicindent}\hspace{\algorithmicindent}\textbf{else}
 \STATE \hspace{\algorithmicindent}\hspace{\algorithmicindent} \hspace{\algorithmicindent}$\mathbb{X}_{u} \gets \mathbb{X}_{u} \cup \{x\}$
 \STATE \hspace{\algorithmicindent}\textbf{end for}
 \STATE \hspace{\algorithmicindent}\textbf{return} $\mathbb{X}_s$, $\mathbb{X}_u$
 \\\textbf{EndFunction}
 \end{algorithmic}
 \end{algorithm}

We now discuss the learning pipeline of the CBF model. 
We write the CBF model as $B_{\theta}: \mathcal{X} \rightarrow \mathbb{R}$.
Given a safe batch $\mathbb{X}_{s}$ and an unsafe batch $\mathbb{X}_{u}$, we optimize $\theta$ by minimizing:
{\small\begin{align}
    &L_{\theta}(\mathbb{X}_{s},\mathbb{X}_{u})\label{eqn:barrier_objective}\\
    &=\Bigg(\frac{1}{|\mathbb{X}_{s}|} \sum_{\small x \in\mathbb{X}_{s}} [-B_{\theta}(x)]_{+}\Bigg) + \Bigg(\frac{1}{|\mathbb{X}_{u}|} \sum_{x \in \mathbb{X}_{u}} [B_{\theta}(x)]_{+}\Bigg)\nonumber\\
     &+ \frac{1}{|\mathbb{X}_{s}|} \sum_{x \in \mathbb{X}_{s}}  \Big[-\bigg\langle \nabla_{x} B_{\theta}(x), \nabla_{x} f(x, \pi_{\Theta}(x))\bigg\rangle\nonumber 
-\alpha \bigg(B_{\theta}(x)\bigg)\Big]_{+}.\nonumber
\end{align}}The first two terms enforce $B_{\theta}(x)$ to take positive values on safe states and negative values on unsafe states, respectively.
The third term optimizes the model to satisfy the Lie derivative condition of CBF in (\ref{equ:barrier_function}).
Unlike prior work~\cite{robey2020learning,yu2023sequential,yu2023learning}, which optimizes the Lie derivative condition over the safe controls from data, we optimize it over the controls generated by the actor $\pi_{\Theta}$.
In fact, optimizing with maximally-safe controls more closely follows the original CBF definition (\ref{equ:barrier_function}) which applies a $\max$ operator over the control space on the Lie derivative.
In Section \ref{sec:simulation_exp}, we show that incorporating the actor
allows for training data collected with
diverse controllers without imposing any performance assumption on them.
Figure \ref{fig:model_component} illustrates the overall training flow of the proposed models.

We present the full procedures in Algorithm \ref{full_alg}.
Early in training, unlabeled data remain unannotated until a sufficient number of iterations have been completed (Line \ref{algline:annotate_condition}).
This is to prevent false model estimations at the outset.
To annotate an unlabeled state $x$, we unroll the dynamics function using the actor's control output to obtain the next state $\bar{x}$ (Line \ref{algline:annotation_unroll}).
We then label $x$ as safe if and only if $\bar{x}$ is deemed safe with respect to the CBF model $B_{\theta}$ \textbf{and} in-distribution with respect to the rejection model $R_{\Phi}$ (Line \ref{algline:annotation_rule}).

 \begin{figure*}[ht]
\centering
\includegraphics[width=0.98\textwidth]{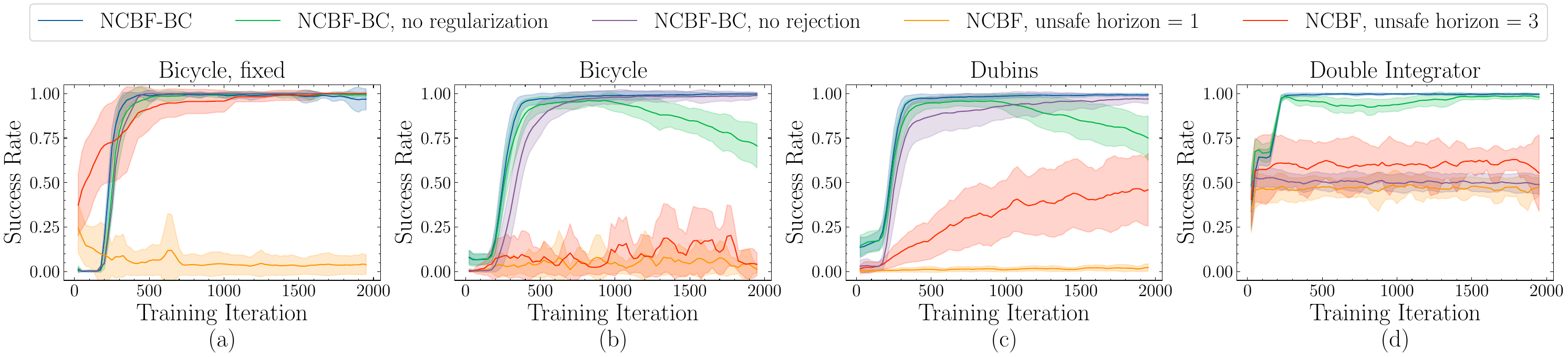} 
\caption{\small Simulation experiments for static obstacle avoidance with different dynamics models of ego-robot.
Evaluation metric is the mean success rate where we follow Algorithm \ref{alg2::derive_control} to derive the controls based on trained models, and perform evaluations over $100$ randomized scenarios.
When collecting the safe trajectories, we employ the potential-field controller(s) with \textbf{(a)} \textit{fixed} and \textbf{(b)-(d)} \textit{randomized} parameters.
}\label{fig:simulation_results}
\end{figure*}
 
\subsection{Optimization Regularization} \label{section:DN}

As the learning objective of the CBF model (\ref{eqn:barrier_objective}) enforces inequality constraints on the estimated landscape,
the training process can suffer from the collapse problem similar to that reported in  self-supervised learning~\cite{jing2021understanding}: 
as training proceeds, the magnitude of the learned landscape may shrink towards near-zero values while still violating the inequality constraints.
This issue is especially pronounced for the offline methods,
as they cannot acquire new data to mitigate the problem unlike the online methods.

To alleviate the collapse issue, 
we optimize (\ref{eqn:barrier_objective}) using the surrogate CBF values $\bar{B}_{\theta}$ defined as follows:
\begin{align}
    \bar{B}_{\theta}(x) &= B_{\theta}(x) / \mathbb{E}_{x \in \bar{\mathbb{X}}} \big[B_{\theta}(x)\big], \label{eqn:dn_denominator}
\end{align}
where $\bar{\mathbb{X}} \subseteq \mathbb{X}_{s}$ is a subset of the safe set sampled in advance.
We do not detach the gradient of the denominator in (\ref{eqn:dn_denominator}) with respect to model parameters $\theta$, which helps to elevate the overall magnitude of barrier landscape whenever it begins to collapse toward zero.

\section{Experiments}
\label{section:experiment}

We evaluate the proposed algorithm in both simulation and real-world experiments.
To derive safety-critical controls from the learned models, we follow Algorithm \ref{alg2::derive_control} which requires a heuristic goal-driven metric to 
rank controls according to task completion progress.


  \begin{algorithm}[b]
 \caption{Control using NCBF-BC} \label{alg2::derive_control}
 \begin{algorithmic}[1]
 \renewcommand{\algorithmicrequire}{\textbf{Input:}}
 \REQUIRE state $x$, CBF model $B_{\theta}$, rejection model $R_{\Phi}$, goal-driven metric $\mathcal{G}: X \veryshortarrow \mathbb{R}$, sample size $N$\\
 \STATE \textit{Sample} control candidates $\textbf{\textit{a}} = [a_1, a_2, ..., a_N]$
 \STATE $\textbf{g} \gets []$
 \FOR{$a_{i}$ in $\textbf{\textit{a}}$}
 \STATE \textit{Unroll} the dynamics $\bar{x}$ $=$ $f(x, a_{i})$ \label{alg2line:unroll}
 \IF{$B_{\theta}(\bar{x} < 0)$ \textbf{or} $R_{\Phi}(\bar{x})$ does not satisfy (\ref{eqn:rejection_condition2})}  \label{alg2line:check_condition}
 \STATE \textit{Remove} $a_{i}$ from $\textbf{\textit{a}}$
 \ELSE
 \STATE \textit{Evaluate} the goal-driven score, $\textbf{g} = \textbf{g} \cup \{\mathcal{G}(\bar{x})\}$
 \ENDIF
 \ENDFOR
 \IF{$\textbf{\textit{a}}$ is now empty}
 \RETURN \textit{Error - no safe control found}  \label{alg2line:error}
 \ENDIF
 \RETURN control candidate from $\textbf{\textit{a}}$ with the maximal score 
 \end{algorithmic}
 \end{algorithm}
 
 \subsection{Simulation Experiments}
\label{sec:simulation_exp}

We focus the simulation evaluations (Figure \ref{fig:simulation_visual}) on the task of static obstacle avoidance.
We generate training trajectories using potential-field controllers with either \textit{fixed} or \textit{randomized} parameters.
A trajectory is considered safe if no collision occurs.
If a trajectory ends in a collision, we add the collision state to the unsafe set, and its preceding segment (of unlabel horizon $\tau$) to the unlabeled set.
In our simulation environments, where the time-step is discretized at $\triangle t = 0.2$ second, we set the unlabel horizon $\tau = 9$.


We consider three different ego-robot dynamics including Double Integrator, the Dubins, and the Bicycle models.
For each dynamics model, the system is of $5$-dimensions consisting of vehicle linear \& angular velocities and yaw angles besides the coordinates. 
All the neural network models are $2$-layer Tanh networks with $128$ hidden neurons per layer.
We use rejection parameter $c = 0.1$,
and perform optimization regularization with subset size $1000$.
We employ the linear mapping $\alpha(x) = \kappa \cdot x$ with $\kappa = 0.1$.
The algorithm runs for $2000$ iterations in total, while the annotation of unlabeled data starts at the $200^{th}$ iteration.
We use the orientation towards the goal as the heuristic goal-driven metric. 
 
\begin{table*}[t]
\centering
\adjustbox{valign=m, width=0.85\textwidth}{
    \begin{tabular}{cccccc}
    \toprule
                 & \begin{tabular}[c]{@{}c@{}}Success Rate \\ (\%)\end{tabular}& \begin{tabular}[c]{@{}c@{}}Mean Path \\ Length (meter)\end{tabular} & \begin{tabular}[c]{@{}c@{}}Mean Completion \\ Time (sec)\end{tabular} & \begin{tabular}[c]{@{}c@{}}Mean Velocity \\ (meter/sec)\end{tabular} & \begin{tabular}[c]{@{}c@{}}Minimal Distance to \\ Obstacles (meter)\end{tabular}\\ 
    \midrule
    NCBF-BC (ours) & \textbf{93.3} & 7.23 & 36.46 & \textbf{0.21} & \textbf{0.578} \\
    NCBF & 46.7 & 7.60 & 35.49 & \textbf{0.21} & 0.610 \\
    PF-1.5m & 80.0 & 7.30 & \textbf{35.06} & 0.20 & 0.668 \\ 
    ROS1-MPC & 80.0 & 7.75 & 44.65 & 0.17 & 0.697  \\
    ROS2-DWB & 86.7 & \textbf{6.68} & 40.26 & 0.17 & 0.677 \\

    \bottomrule
    \end{tabular}
}
\caption{\small Real-world experiments for dynamic obstacle avoidance over $30$ runs.
Statistics are computed over the successful runs only.
}
\label{table:real_world_dynamic}
\end{table*}

\begin{figure}[b]
\centering
\includegraphics[width=\linewidth]{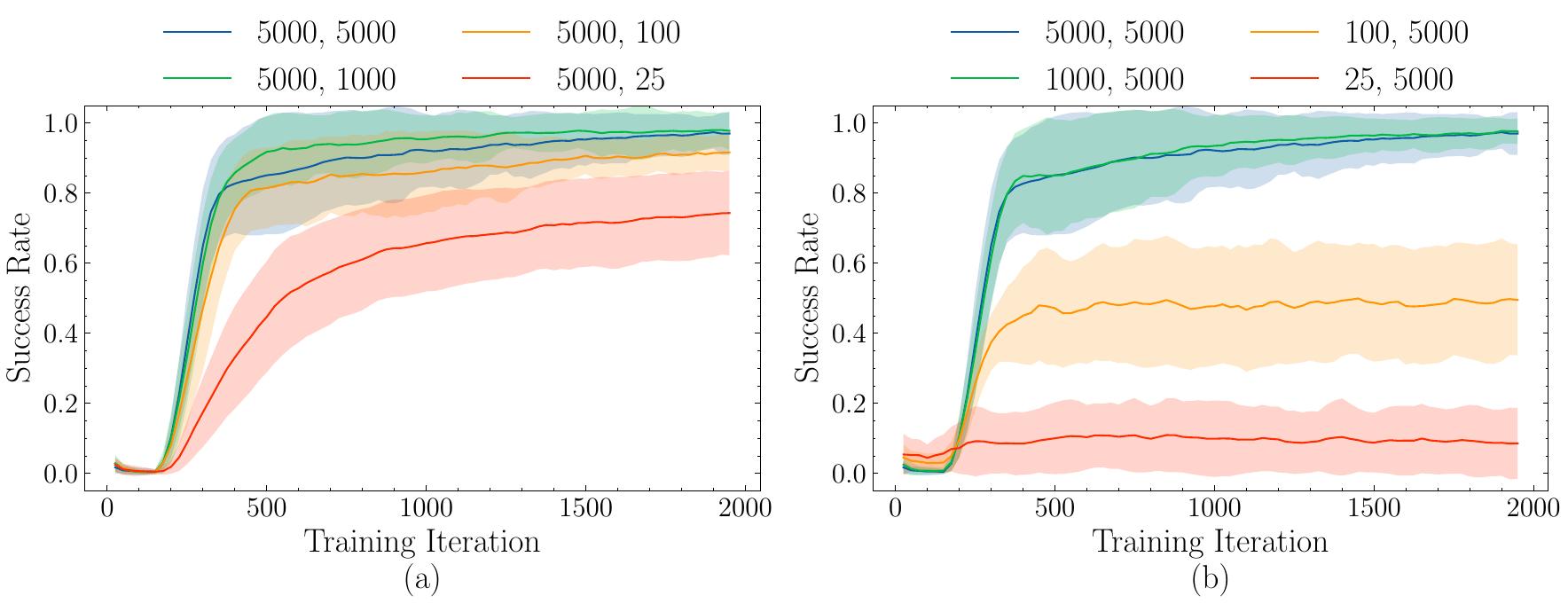}
\caption{\small Ablation experiments on the ratio between training labeled and unlabeled set sizes for Bicycle.
The two numbers in each label are the sizes of labeled and unlabeled demonstration sets, respectively.
For instance, the blue curve in (a) refers to the setting with $5000$ labeled safe \& unsafe states, and $5000$ unlabeled states.
} \label{fig:simulation_results3}
\end{figure}

 
\noindent{\textbf{Comparisons with Baseline Methods}.}
Figure \ref{fig:simulation_results} demonstrates the experiment results.
The baseline is the standard method for learning CBFs as in \cite{dawson2022safe,robey2020learning,qin2022quantifying,yu2023sequential,yu2023learning}, denoted by \textit{Neural CBF} (NCBF).
The \textit{Unsafe horizon} defines the number of states near the end of failure trajectories which we label as unsafe, but only when training the baseline.

First, we show that the baseline underperforms when training trajectories are generated by multiple controller polices.
When there are multiple controls provided at a state, 
it is incorrect to optimize the Lie derivative condition (\ref{equ:barrier_function}) of CBF over all the provided safe controls.
Instead, we should optimize the condition only along the maximally-safe control, while those less conservative controls can be allowed to violate the inequality constraint in (\ref{equ:barrier_function}). 
In Figure \ref{fig:simulation_results}(b)-(d), we show that our method can handle training sets collected by a diverse range of sub-optimal controllers.

Across all experiments, the proposed method incorporating both regularization and rejection-based annotation performs the best in terms of learning rates and training stability.
For the Bicycle and Dubins models, the regularization technique effectively prevents collapse, thereby avoiding divergence during training. 
The Double Integrator does not exhibit collapse issues, regardless of whether regularization is applied.
Because it has simpler dynamics, the optimization for satisfying the CBF conditions converges before collapse can occur.
Meanwhile, disabling the rejection-based annotation noticeably slows convergence.
In particular, for the Double Integrator, the CBF quickly overfits to the available labeled data, 
weakening the effectiveness of the annotation process that relies only on the learned CBF scores.


\noindent\textbf{Ratio of Labeled Data}. We conduct ablation experiments on the Bicycle model to investigate how the ratio of training set size impacts the proposed method.
In Fig. \ref{fig:simulation_results3}(a), we vary the size of unlabeled set while fixing the labeled size,
showing that the learning can be quickly stimulated even with small amounts of unlabeled data.
This is because the labeled states that are certainly safe may be derived from conservative policies whose safety rules deviate from the optimal safety boundary.
Meanwhile, the unlabeled trajectories with uncertain safety often involve aggressive controls that may exceed the optimal safety boundary, and thus carry more useful information.
In Fig. \ref{fig:simulation_results3}(b), we fix the unlabeled size while varying the size of labeled set.
With sufficient unlabeled data provided, 
there appears to be a threshold to the labeled size beyond which the learning rates become indifferent. 

\begin{figure}[b]
    \centering
    \adjustbox{valign=b, width=0.2\textwidth}{
    \subfigure{\includegraphics{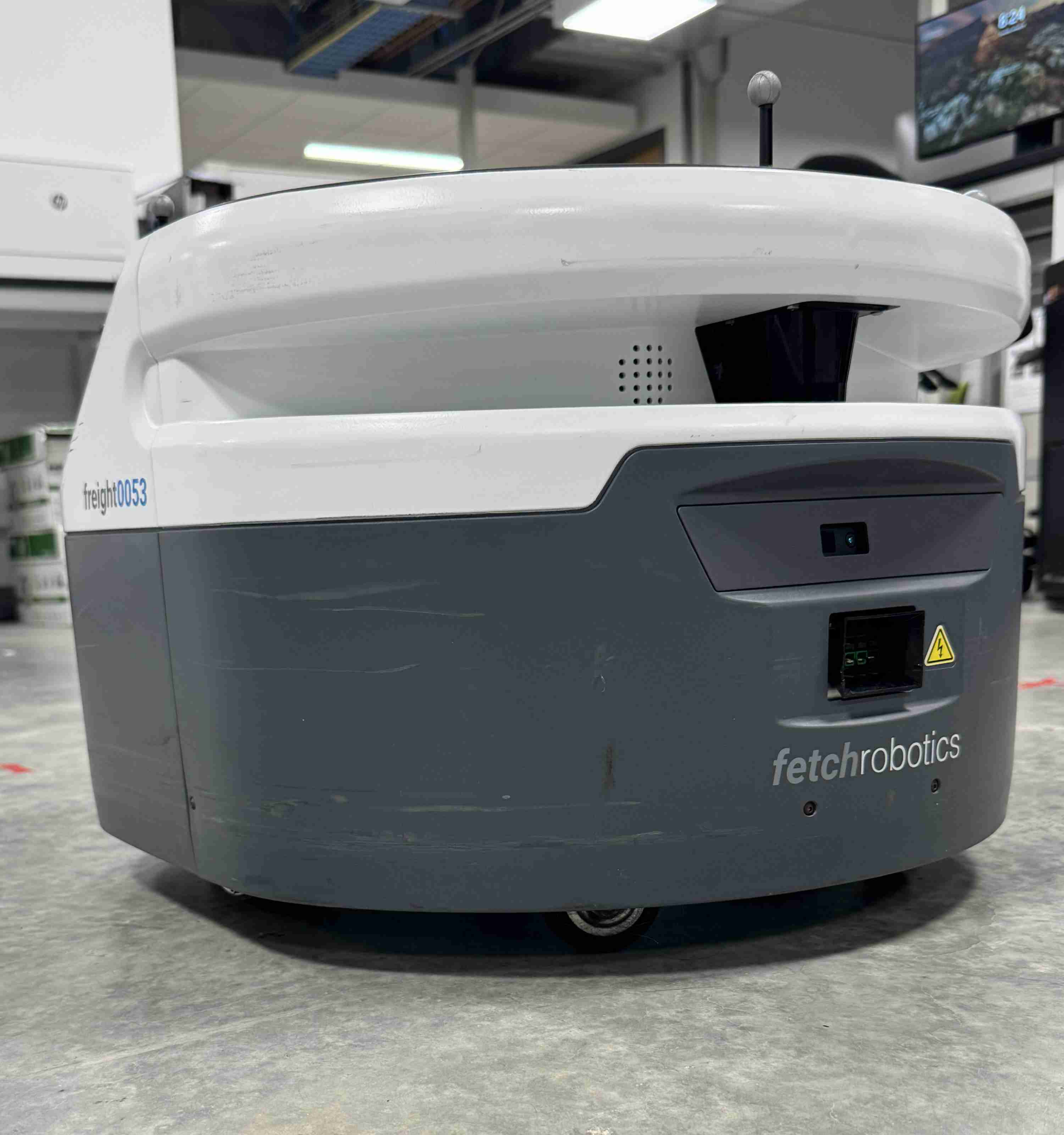}}}
    \adjustbox{valign=b, width=0.45\linewidth}{
    \subfigure{\includegraphics{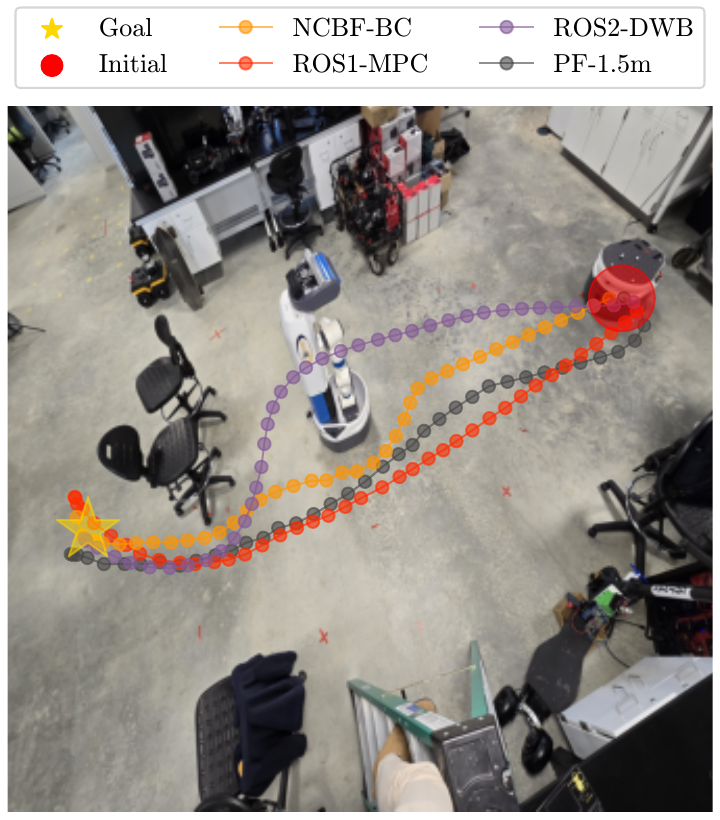}}}
    \caption{\small \textbf{Left}: Freight robot. 
    \textbf{Right}: Trajectories generated by different controllers. 
    }\label{fig:simulation_results2}
\end{figure}



\subsection{Hardware Experiments}\label{sec:real_world_exp}

\begin{figure}[t]
    \centering
    \adjustbox{width=0.45\textwidth}{
        \includegraphics{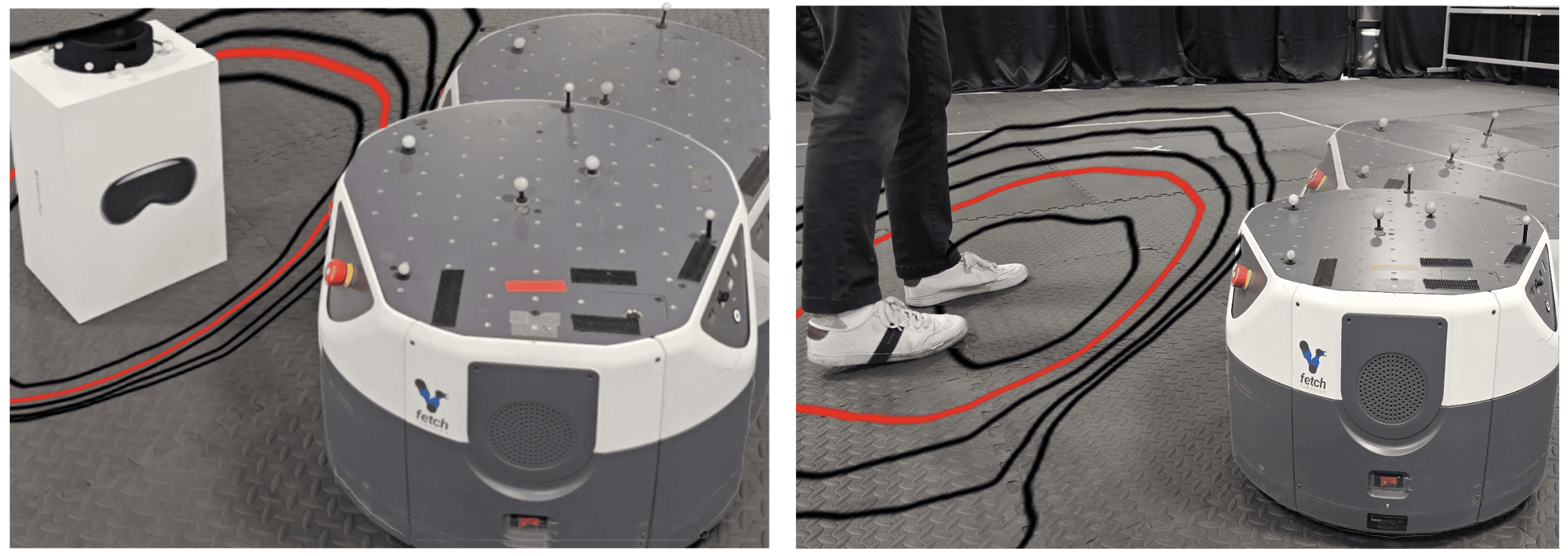}
    }
    \caption{\small Visualization of the learned CBF landscapes.
    Note that the CBF model trained for dynamic obstacle avoidance exhibits a wider gap between level sets, 
    which reflects the need to initiate collision avoidance further from dynamic obstacles compared to static ones.
    }
    \label{fig:realworld_visual}
\end{figure}

In this section, we discuss hardware experiments on dynamic obstacle avoidance (Figure \ref{fig:realworld_visual}).
The experiments on static obstacles (Figure \ref{fig:simulation_results2} Right) are showcased in supplementary video but are not discussed in the paper.
The platform utilized in our experiments is the Freight (Figure \ref{fig:simulation_results2} Left), a research variant from Fetch Robotics. 
We cap the velocity of the robot at 0.22 m/s. 
To train our model, we collected $40$-minutes of demonstrations by manually driving the robot around pedestrians,
deliberately splitting the data into roughly $15$-minutes of successful and $25$-minutes of failure trajectories.
For applying the proposed method, we discretize the time-step to be $\Delta t = 0.15$ second, and employ unlabel horizon $\tau = 9$.

The system state space for dynamic obstacle avoidance
is $11$-dimensional,
consisting of robot coordinates, yaw and velocity information, and $3$-step past state history of individual pedestrian.
All the neural models are $2$-layer Tanh networks with $256$ hidden neurons per layer. 
We perform the learning for $5000$ iterations, initiating the annotation steps over unlabeled data at the $500^{th}$ iteration.
All other training parameters match those used in simulation experiments.
When optimizing (\ref{eqn:barrier_objective}) to enforce the Lie derivative condition of CBF, we only leverage the derivative of ego-robot dynamics, while taking from data the pedestrian movements at the future timestamps.

Table $\ref{table:real_world_dynamic}$ presents the quantitative results.
Besides \textit{NCBF}, we include one potential-field controller using a repulsive range of $1.5$ meter. 
We further compare against both the ROS1 MoveBase (MPC-based) navigation stack and the ROS2 Nav2 stack.
Offline Deep RL algorithms cannot be directly applied,
as the reward labels that can accurately reproduce manual controls (via RL) are typically unavailable in real-world data.
First, the proposed method achieves the highest success rate.
The failures with our method are always due to unfamiliar pedestrian movements that deviate from the training data.
Second, we show that the proposed method completes the scenarios with the highest mean velocity, while maintaining the lowest distance to the pedestrians without violating safety.
This showcases the robustness of the learned safety boundary which allows us to select the controls that are performative, or even aggressive, yet safe.
Third, the performance of potential-field controller degrades when there involve more pedestrians surrounding the robot.
Oscillation behaviors are observed with potential-field controller in our experiments.
Last,
the NCBF baseline shows sub-optimal performance, producing strange looping behaviors and frequently taking unnecessarily long paths when pedestrians are present.
Since data were collected via manual control, states could be reached with controls of varying levels of conservativeness.
Consequently, the baseline's training objective forces the CBF to accommodate the most conservative control among the provided examples.
Moreover, the NCBF baseline under-utilize failure trajectories, especially the uncertain states preceding the collisions, thereby limiting the amount of data it can effectively leverage.


\section{Conclusion}
\label{section:conclusion}

This paper presents a novel learning-based approach for constructing neural control barriers from offline demonstrations.
Experiment results show that the proposed algorithm outperforms the existing methods for offline data-driven CBFs construction.
The results also highlight the benefits of utilizing unlabeled demonstrations to further refine the learning of control barriers
Future directions include extending the proposed method to higher-dimensional systems, such as those incorporating Lidar sensory data or images.

\newpage
\bibliographystyle{ieeetr}
\bibliography{references}

\end{document}